\documentclass[sn-aps]{sn-jnl}

\usepackage{graphicx}%
\usepackage{multirow}%
\usepackage{amsmath,amssymb,amsfonts}%
\usepackage{amsthm}%
\usepackage{mathrsfs}%
\usepackage[title]{appendix}%
\usepackage{xcolor}%
\usepackage{textcomp}%
\usepackage{manyfoot}%
\usepackage{booktabs}%
\usepackage{algorithm}%
\usepackage{algorithmicx}%
\usepackage{algpseudocode}%
\usepackage{listings}%
\usepackage{siunitx}%
\usepackage{natbib}

\usepackage[final]{changes}
\usepackage[usenames,dvipsnames]{xcolor} 
\definechangesauthor[name={Liu}, color=red]{PS}

\begin{document}

\listofchanges 

\title[Article Title]{Fabrication and characterization of AlMn alloy superconducting films for 0$\nu\beta\beta$ experiments}


\author[1]{\fnm{Zhouhui} \sur{Liu}}\email{liuzh@ihep.ac.cn}
\author*[1,2]{\fnm{Yifei} \sur{Zhang}}\email{zhangyf@ihep.ac.cn}
\author[1]{\fnm{Yu} \sur{Xu}}\email{xuyu97@ihep.ac.cn}
\author[3]{\fnm{Mengxian} \sur{Zhang}}\email{zhangmengxuan@ihep.ac.cn}
\author[4]{\fnm{Qing} \sur{Yu}}\email{q-yu@mail.tsinghua.edu.cn}
\author[1]{\fnm{Xufang} \sur{Li}}\email{lixufang@ihep.ac.cn}
\author[1]{\fnm{He} \sur{Gao}}\email{hgao@ihep.ac.cn}
\author[1,2]{\fnm{Zhengwei} \sur{Li}}\email{lizw@ihep.ac.cn}
\author[1,2]{\fnm{Daikang} \sur{Yan}}\email{yandk@ihep.ac.cn}
\author[1,2]{\fnm{Shibo} \sur{Shu}}\email{shusb@ihep.ac.cn}
\author[1,2]{\fnm{Yongjie} \sur{Zhang}}\email{zhangyj@ihep.ac.cn}
\author[1]{\fnm{Xuefeng} \sur{Lu}}\email{luxf@ihep.ac.cn}
\author[5]{\fnm{Yu} \sur{Wang}}\email{986215512@qq.com}
\author[5]{\fnm{Jianjie} \sur{Zhang}}\email{zhaofacangwu@126.com}
\author*[5]{\fnm{Yuanyuan} \sur{Liu}}\email{yyliu@bnu.edu.cn}
\author*[1,2]{\fnm{Congzhan} \sur{Liu}}\email{liucz@ihep.ac.cn}

\affil*[1]{\orgdiv{State Key Laboratory of Particle Astrophysics, Institute of High Energy Physics}, \orgname{Chinese Academy of Sciences}, \orgaddress{\city{Beijing 100049}, \country{China}}}

\affil[2]{\orgdiv{School of Physical Science}, \orgname{University of Chinese Academy of Sciences}, \orgaddress{\city{Beijing 100049}, \country{China}}}

\affil[3]{\orgdiv{School of Mechanical and Elecrical Engineering}, \orgname{China University of Mining and Technology}, \orgaddress{\city{Xuzhou 221116}, \country{China}}}

\affil[4]{\orgdiv{Department of Astronomy}, \orgname{Tsinghua University}, \orgaddress{ \city{Beijing 100084}, \country{China}}}

\affil[5]{\orgdiv{College of Physics and Astronomy}, \orgname{Beijing Normal University}, \orgaddress{\city{Beijing 100875}, \country{China}}}


\abstract{Neutrinoless double-beta decay (0$\nu\beta\beta$) experiments constitute a pivotal probe for elucidating the characteristics of neutrinos and further discovering new physics. Compared to the neutron transmutation doped germanium thermistors (NTD-Ge) used in 0$\nu\beta\beta$ experiments such as CUORE, transition edge sensors (TES) theoretically have a relatively faster response time and higher energy resolution. These make TES detectors good choice for next generation 0$\nu\beta\beta$ experiments. In this paper, AlMn alloy superconducting films, the main components of TES, were prepared and studied. The relationship between critical temperature ($T_c$) and annealing temperature was established, and the impact of magnetic field on $T_c$ was tested. The experimental results demonstrate that the $T_c$ of AlMn film can be tuned in the required range of 10 - 20 mK by using the above methods, which is a key step for the application of AlMn TES in 0$\nu\beta\beta$ experiment. In the test range, the $T_c$ of AlMn film is sensitive to out-of-plane magnetic field but not to the in-plane magnetic field. Furthermore, we find that a higher annealing temperature results in a more uniform distribution of Mn ions in depth, which opens a new avenue for elucidating the underlying mechanism for tuning $T_c$.}


\maketitle

\section{Introduction}\label{sec1}

The neutrinoless double-beta decay (0$\nu\beta\beta$) search is one of the most important studies in current physics. It could prove whether the neutrino is a Majorana fermion (i.e., its own antiparticle), whether lepton number is conserved, and determine the mass hierarchy of neutrinos \cite{RivNuovoCim2024, RevModPhys.95.025002}. Currently, there are many running projects, such as CUORE, GERDA, KamLAND-Zen, AMoRE, PandaX and so on, which are in development \cite{doi:10.1142/S0217751X25500502, 10.1093/ptep/ptad038, 2024Projected, 2025Light, 2025PandaX, Aalbers_2025}. CUPID, upgrade from CUORE, \replaced[id=PS]{will use}{uses} the $^{100}$Mo isotope with a Q$_{\beta\beta}$ of 3.034 MeV as the detection target to obtain a relatively lower natural background in the region of interest \cite{FIORINI1998309, 2025CUPID}. It inherits the low temperature bolometric technology of CUORE and adds another bolometer to detect scintillation light from the main crystal to further discriminate the alpha particle background \added[id=PS]{- the viability of this technology was successfully demonstrated in CUPID-Mo experiment }\cite{2022Final, PhysRevLett.126.181802}. These make CUPID one of the most competitive \added[id=PS]{future }experiments in this area. In the development of the CUPID project, several potential light detector technologies were considered, including MKID, TES, and NTD-Ge \cite{armatol2022cupid1t, thecupidinterestgroup2019cupidprecdr, PhysRevApplied.20.064017}. TES sensors have the characteristics of fast response time and high energy resolution, which can be readout multiplexing through SQUID amplifier \cite{instruments8040047, 2023Transition}. They are considered to be a promising option for 0$\nu\beta\beta$ \replaced[id=PS]{detectors}{experiment}. Inspired by the CUPID detector concept, a CUPID-like scintillating crystal bolometer based on TES technology is currently being developed for the 0$\nu\beta\beta$ experiment which will be scheduled to be carried out at China JinPing underground Laboratory (CJPL) \cite{armatol2022cupid1t, 2022Cosmogenic}.

There have been many reports on the study of superconducting film materials, including single-element films, double-layer films utilizing the proximity effect of superconductivity, and alloy films doped with magnetic impurities \cite{10.1063/1.1789575}. Among them, the aluminum doped with manganese (AlMn) alloy superconducting film has been widely used in astronomical and astrophysical instruments due to its easy tuning of critical temperature ($T_c$)\cite{2004Dilute}. However, there has been little research on AlMn alloy films at extremely low temperatures, especially in the application of 0$\nu\beta\beta$ experiments. 

In this paper, several researches have been conducted with the goal of applying AlMn alloy films in 0$\nu\beta\beta$ experiment. It was measured how preparation parameters affect the sputtering rate of AlMn alloy films. Monolayer AlMn alloy films with different thickness and Mn concentration are prepared. Based on the annealing method, the relationship between $T_c$ and annealing temperature of AlMn alloy films is plotted, from which the optimization fabrication parameters and annealing temperature of AlMn alloy films suitable for 0$\nu\beta\beta$ experiment are found. In addition, the effects of magnetic field and current on the $T_c$ are also studied, as well as the distribution of Mn impurity in AlMn films. 

\section{Fabrication of AlMn films}\label{sec2}


\begin{figure}[htbp]
    \centering
    \includegraphics[width=0.93\textwidth]{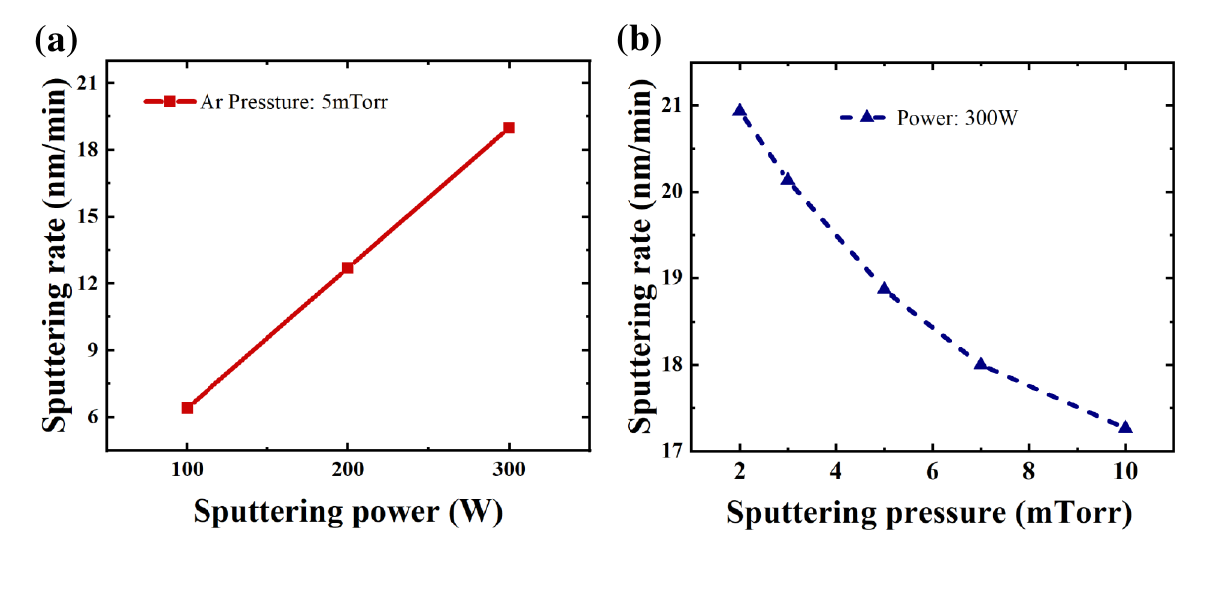}
    \caption{Influence of sputtering power and Ar pressure on sputtering rate. (a) The red data points indicate the relationship between sputtering rate and sputtering power. (b) The blue data points indicate the relationship between sputtering rate and sputtering pressure.}
    \label{fig:1}
\end{figure}

Manganese impurities can break the Cooper pairs in aluminum alloy films, thereby significantly reducing the critical temperature $T_c$ of the alloy films compared to pure aluminum films \cite{2004Dilute}. According to previous studies, two AlMn alloy targets with Mn impurity of 1800 ppm and 2000 ppm, respectively, were utilized to fabricate superconducting films with $T_c$ below 20 mK. These targets were placed in a DC magnetron sputtering deposition system, which has four fixed target bases and can ensure an ultra-high base vacuum of $5\times 10^{\text{-}9}$ Torr. With the target–substrate distance fixed at 153 mm, the effects of sputtering power and argon gas pressure on the sputtering rate during the preparation of AlMn alloy films were studied. These calibrated values are used to set the sputtering time, ensuring accurate thickness control for every subsequent deposition. In each study, the thickness of films was measured using a step profiler. Fig.1(a) shows that the sputtering rate ($V_s$) has a linear relationship with the sputtering power ($P_s$), and the relation is $V_s = 0.063P_s + 0.1$, as indicated by the red fitted line. In contrast to power, the sputtering rate is relatively insensitive to argon pressure. Specifically, when the argon pressure is increased from 2 mTorr to 10 mTorr, the sputtering rate only decreases by approximately 20 percent, as shown in Fig1.(b). During the film preparation process, the sputtering rate can be effectively controlled by adjusting the sputtering power.

\section{Characterization of AlMn films}\label{sec3}

The resistance of the superconducting film varies with the temperature from a normal state to a superconducting state. The superconducting films have two important parameters, the critical temperature $T_c$ and superconducting transition width $\Delta T_c$, which play a key role in the performance of superconducting detectors. The normal resistance of the superconducting film is denoted as $R_n$, $T_c$ is taken to be the temperature value at 0.5$R_n$, and the $\Delta T_c$ represents the temperature interval between 0.9$R_n$ and 0.1$R_n$. The characterization of AlMn alloy films deposited by DC magnetron sputtering is assessed using a Bluefors LD250 dilution refrigerator (DR) with the four-terminal method \cite{Miccoli_2015}, as shown in Fig.2(a). \added[id=PS]{In the lower-right inset of Fig. 2(a), the sample sizes of measured AlMn alloy films are approximately 5 mm $\times$ 5 mm, and the two outer contacts inject the current while the two inner pair measure the voltage drop, allowing us to record R–T curves and extract $T_c$. }The superconducting transition curve R-T of AlMn alloy films can be obtained, as shown in Fig.2(b).

\begin{figure}[htbp]
    \centering
    \includegraphics[width=0.93\textwidth]{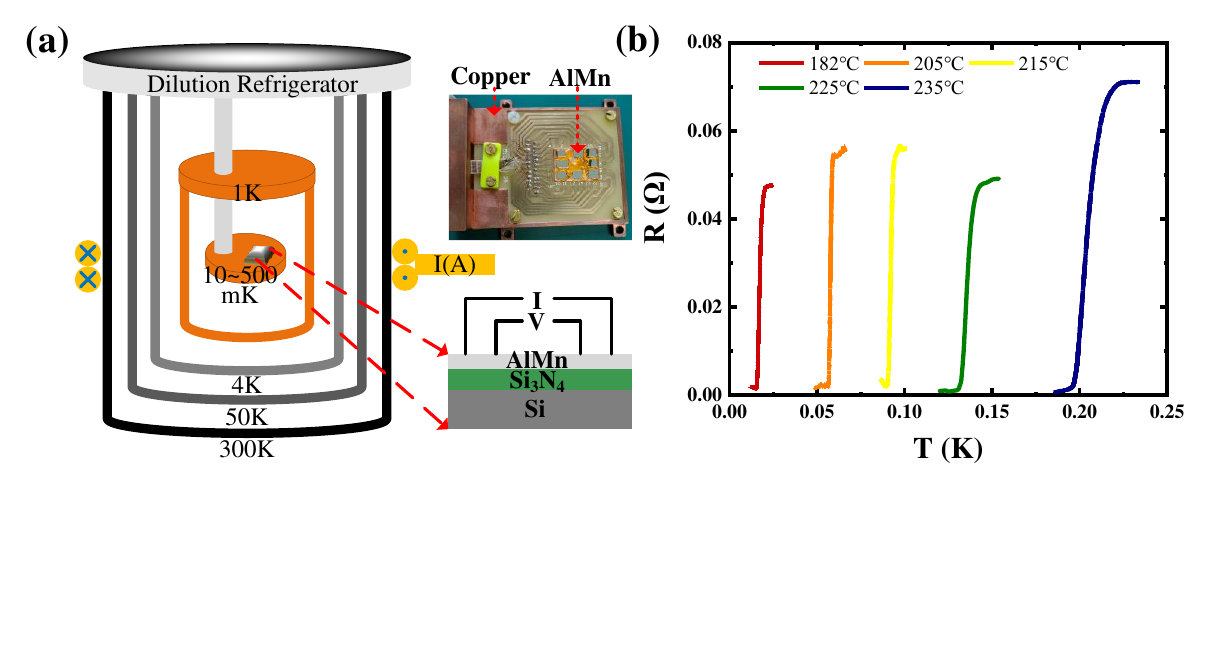}
    \setlength{\abovecaptionskip}{-1cm} 
    \caption{(a) R-T of films are characterized by the four-terminal method in a dilution refrigerator. \added[id=PS]{Upper-right inset: schematic of the thin-film sample and sample holder used for the measurements. }(b) Example of R versus T measured by the four-terminal method with a DC current source around film superconducting transition. \added[id=PS]{The horizontal axis represents the measurement temperature, and the vertical axis shows the film resistance measured by the four-terminal method; the five R–T curves correspond to different annealing temperatures.}}
    \label{fig:2}
\end{figure}

The influence of Mn concentration, film thickness, and annealing temperature on $T_c$ of AlMn alloy films is studied. The 2000 ppm AlMn alloy films are explored with different thicknesses and annealing temperatures. As shown in Fig.3(a), the $T_c$ of AlMn alloy films initially decreases and then increases with rising annealing temperature. When the annealing temperature is below approximately $\SI{235}{\degreeCelsius}$, the $T_c$ of AlMn alloy films decreases as the increase of film thickness at the given annealing temperature. However, when the annealing temperature exceeds $\SI{235}{\degreeCelsius}$, the opposite trend is observed. It should be noted that the parameters within the gray dashed circle of Fig.3(a) correspond to films whose $T_c$' below the DR's base temperature of 7 mK, no superconducting transition could be detected within the present experimental limit. The $\Delta T_c$ of the 2000 ppm AlMn alloy films, which is shown in Fig.3(b), remains relatively stable about 2 mK at first and then increases with increasing annealing temperature, showing minimal dependence on film thickness. When the annealing temperature surpasses $\SI{200}{\degreeCelsius}$, $\Delta T_c$ increases significantly. These observations are crucial for selecting doping concentrations and annealing temperatures in TES detector fabrication. A smaller $\Delta T_c$ corresponds to a steep transition width and a larger temperature-sensitivity coefficient $\alpha \propto \partial R/\partial T$, which in turn yields a higher temperature sensitivity\cite{pyle2015optimizeddesignslowtemperature}. During the fabrication of the detector, the sample temperature generally does not exceed $\SI{120}{\degreeCelsius}$. Consequently, for applications demanding high sensitivity, AlMn films should be annealed in the 120–200 °C range, where $\Delta T_c$ is minimized without violating standard fabrication temperature limits. Further analysis of the R-T curve of the film in Fig.3(a) shows that the $T_c$ of AlMn alloy films is approximately linear with the annealing temperature in the range from $\SI{180}{\degreeCelsius}$ to $\SI{250}{\degreeCelsius}$, which aids in finding the target with low $T_c$.

\begin{figure}[htbp]
    \centering
    \includegraphics[width=0.93\textwidth]{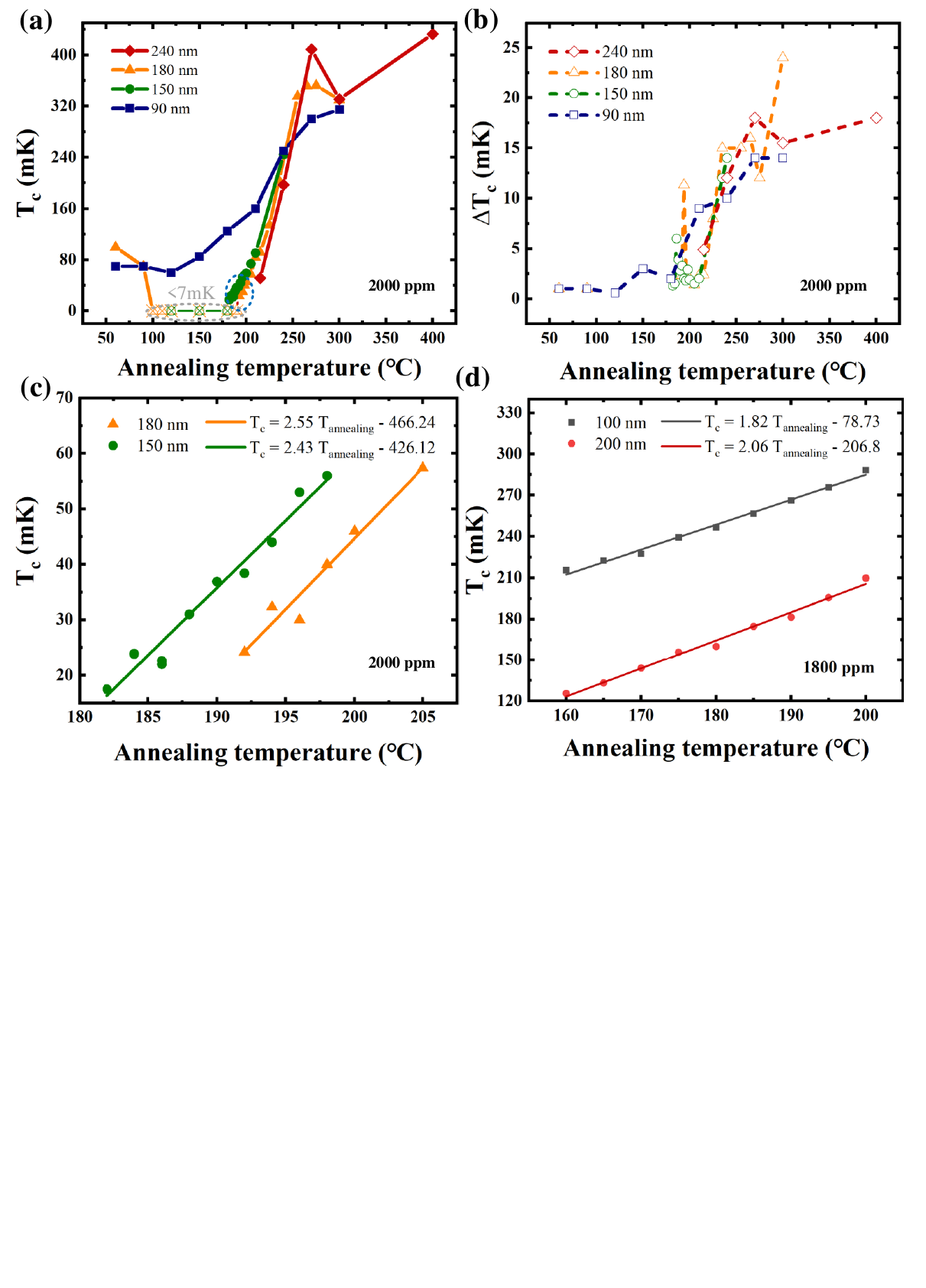}
    \setlength{\abovecaptionskip}{-6cm} 
    \caption{(a) The change of $T_c$ of 2000 ppm AlMn alloy film with film thickness and annealing temperature. (b) The change of $\Delta T_c$ of AlMn alloy film with film thickness and annealing temperature. (c) The enlarged view of the data corresponding to the annealing temperature of 182 to $\SI{205}{\degreeCelsius}$ in Fig3.(a), with linear fitting of the data. (d) The $T_c$ of 1800 ppm AlMn alloy films baked above $\SI{150}{\degreeCelsius}$ presents a linear relationship with the annealing temperature, and the slope decreases with the decrease of Mn doping concentration.}
    \label{fig:3}
\end{figure}
Fig.3 (c) provides a detailed view of the data within the blue dashed area in Fig.3(a). Linear fitting is performed on the AlMn alloy films with thickness of 150 nm and 180 nm specifically within the annealing temperature range of $\SI{182}{\degreeCelsius}$ to $\SI{205}{\degreeCelsius}$. \deleted[id=PS]{The green line corresponds to the equation $T_c$ = 2.43 $T_{annealing}$ - 426.12, and the yellow line corresponds to $T_c$ = 2.55 $T_{annealing}$ - 466.24.} The two lines exhibit very similar slopes. Based on this trend, the AlMn alloy film with $T_c$ of 17 mK is measured, whose annealing temperature is $\SI{182}{\degreeCelsius}$ and thickness is 150nm. This is invaluable for fabricating TES detectors tailored for the 0$\nu\beta\beta$ experiment.

Furthermore, the 1800 ppm AlMn alloy films with thicknesses of 100 nm and 200 nm are fabricated and baked at temperatures ranging from $\SI{160}{\degreeCelsius}$ to $\SI{200}{\degreeCelsius}$, with the step increment of $\SI{5}{\degreeCelsius}$. The relationship between $T_c$ and annealing temperature of those 1800 ppm AlMn alloy films is shown in Fig.3(d). It can be found that there is also a linear relationship between the $T_c$ and annealing temperature. The linear fitting equations of $T_c$ and annealing temperature for the 100 nm (black line) and 200 nm (red line) AlMn films in the Fig.3(d) are $T_c$ = 1.82 $T_{annealing}$ - 78.73 and $T_c$ = 2.06 $T_{annealing}$ - 206.8, respectively. These equations reveal that the slopes of the two lines are similar. By following the linear trend of the curve, it can be inferred that 1800 ppm AlMn alloy films can be annealed at lower temperatures to achieve a very low $T_c$ of 10 - 20 mK\added[id=PS]{. Data for AlMn films with other Mn concentration and additional film thicknesses have been reported in earlier work}\cite{Yu_2021}. When comparing Fig. 3(c) and (d), it is evident that the $T_c$ of AlMn films increases with the decrease of Mn content within the AlMn target. Simultaneously, the slope of the fitting line also increases. This indicates that the influence of $T_c$ on annealing temperature can be reduced with the increase of Mn doping concentration.

\begin{figure}[htbp]
    \centering
    \includegraphics[width=0.93\textwidth]{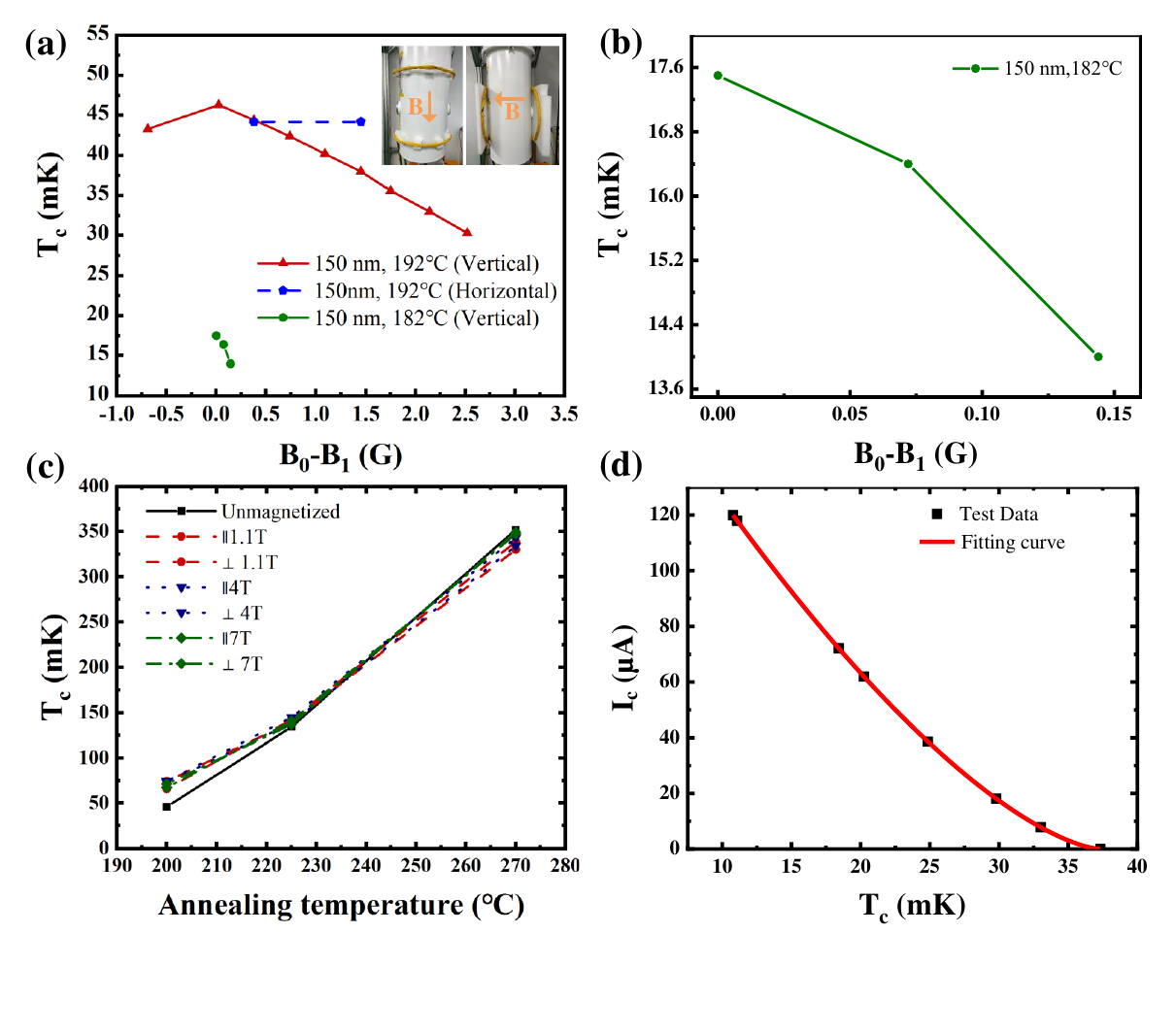}
    \setlength{\abovecaptionskip}{-1cm} 
    \caption{Relationship between $T_c$ of 2000 ppm AlMn alloy films and external magnetic field, as well as magnetization of strong magnetic field. (a) and (b) The variation of $T_c$ of 150 nm AlMn alloy film with external magnetic field. (c) The $T_c$ of magnetized 2000 ppm AlMn alloy films exposed to different magnetic fields is compared with that of unmagnetized films. (d) The $T_c$ and critical current $I_c$ of AlMn alloy films satisfy the function $I_c(T_c)=I_{c0}(1-T_c/T_{c0})^3{^{{^/2}}}$ with $I_{c0}=200 \mu{A}$ and $T_{c0}$=37.3 mK, which is consistent with the Ginzburg-Landau (GL) theory.}
    \label{fig:4}
\end{figure}

It is known that superconducting devices are sensitive to magnetic fields. When a magnetic field is applied to the superconducting film in the TES detector, it can cause a shift in the $T_c$, which can affect the sensitivity of the device, according to the formula of the mean energy fluctuation
$\sigma^2 E \propto k_B T^2 C/\alpha$ \cite{pyle2015optimizeddesignslowtemperature}. Magnetic sources that can interfere with TES include natural geomagnetic fields and man-made fields generated by nearby
instrumentation \cite{2017Magnetic}. To investigate the effect of magnetic fields on the $T_c$ of AlMn alloy films, two self-made Helmholtz coils were put on the outside of the refrigerator, as shown in the inner panel of Fig.4(a). It was calibrated by a MEDA FVM-400 vector magnetometer. Here, $B_0$ represents the geomagnetic field, and $B_1$ denotes the magnetic field generated by coils. As shown in Fig.4 (a), the vertical magnetic field significantly impacts the $T_c$ of AlMn alloy filmsat a rate of approximately - 6.4 mK/G, while the horizontal magnetic field has no significant effect within the relevant range. Moreover, AlMn alloy films with a lower $T_c$  more susceptible to be affected by the magnetic field. With an increase of about 0.14 G in the magnetic field, the $T_c$ of AlMn alloy film will decrease remarkably from 17.6 mK down to 14.0 mK, which is the lower $T_c$ our present setup can reliably test, as shown in Fig.4 (b). Therefore, when using AlMn alloy film as a TES in 0$\nu\beta\beta$ experiment, magnetic shielding is necessary to mitigate this effect. 

Furtherly, three 2000 ppm AlMn alloy films baked at different temperatures are selected, and exposed them at three distinct magnetic fields of 1.1 T, 4 T, and 7 T for 10 minutes at room temperature. Both parallel and vertical magnetization directions were considered. The $T_c$ of those films was extracted from R-T curves and shown in Fig.4 (c). There is a little difference between unmagnetized films and magnetized films. The magnetic field in the test range hardly affects $T_c$ of AlMn alloy films baked at the same temperature. These results indicate that room-temperature magnetization has minimal effect on the properties of AlMn films.

Furthermore, an AlMn film with the thickness 150 nm was selected to evaluate the impact of current on $T_c$. During the test, a specific current denoted as $I_c$ was applied. The sink temperature slowly rose from 10 mK to a higher temperature until the superconductivity was broken. Fig.4 (d) plots the result for a 2000 ppm AlMn alloy film baked at $\SI{190}{\degreeCelsius}$. The experimental data can be \deleted[id=PS]{accurately} fitted using the function $I_c(T_c)=I_{c0}(1-T_c/T_{c0})^3{^{{^/2}}}$, in which $I_{c0}=200 \mu{A}$, $T_{c0}$=37.3 mK. It is consistent with the Ginzburg-Landau theory \cite{2012A,10.1063/1.367153}.

\begin{figure}[htbp]
    \centering
    \includegraphics[width=0.93\textwidth]{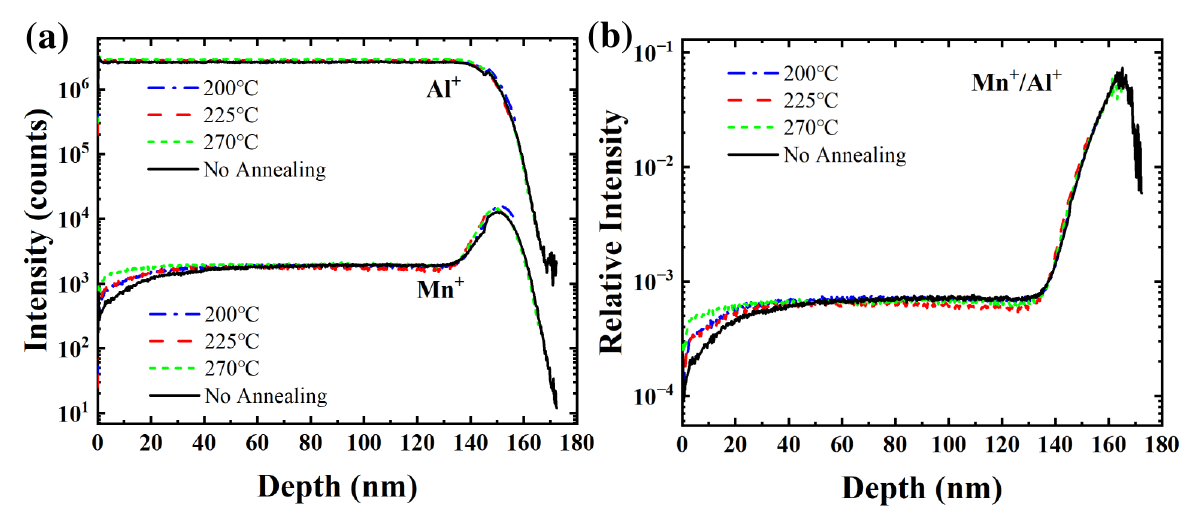}
    \caption{The measured distribution of $Al^+$ and $Mn^+$ content in depth for \replaced[id=PS]{annealed and one as-deposited}{baked and unbaked} AlMn alloy films. (a) The absolute numbers of ions at different depths. (b) The ratio of the number of $Al^+$ to $Mn^+$.}
    \label{fig:5}
\end{figure}

The exact physical mechanism governing the variation of the $T_c$ of AlMn alloy films with annealing temperature remains unknown \cite{2016AlMn}. Lv et al. proposed that annealing might resume the exchange coupling among Mn impurities, thereby enhancing the superconducting temperature of AlMn alloy films \cite{2021Realization}. Another research team reported using extended X-ray absorption fine spectroscopy (EXAFS) and found that the Mn clustering increases with increasing annealing temperature \cite{2023APS..MARS27008B}. Here, Time of Flight Secondary Ion Mass Spectrometry (TOF-SIMS) was employed to characterize three thermally annealed and one as-deposited AlMn films. Fig.5 shows that significant Mn ion accumulation occurs at the interface between the AlMn alloy film and silicon nitride. As the annealing temperature rises, the intensity of Mn ions near the surface of the AlMn film increases, while the number of Mn ions at the mid-depth declines gradually. These results indicate that elevated annealing temperatures enhance the homogeneity of Mn ion distribution, potentially contributing to the modulation of the $T_c$ of AlMn films. However, further study with more experimental data is required to confirm this hypothesis.

\section{Conclusion}\label{sec4}

We fabricated and characterized AlMn alloy films, which will serve as the core superconducting films of TES detectors used in 0$\nu\beta\beta$ experiment at CJPL. The influence of sputtering power and Ar pressure on the sputtering rate of AlMn films was studied. Monolayer AlMn films were prepared with different thicknesses, Mn doping concentrations, and annealing temperatures. The R-T curves of these AlMn films were measured using the four-terminal method, from which the $T_c$ and $\Delta T_c$ of the films were extracted. For AlMn targets with various Mn doping concentrations, the sputtering rate and power exhibit a linear relationship. The $T_c$ of the AlMn film increases with the decrease of Mn content in AlMn targets at the same annealing temperature. Meanwhile, the effect of annealing temperature on $T_c$ is enhanced with higher Mn doping concentrations. \replaced[id=PS]{By systematically analyzing the relationship between the AlMn alloy films and annealing temperature, the fabrication window can be rapidly pinpointed in which $T_c$ falls between 10 and 20 mK. Fine-tuning an external magnetic field further lowers the minimum $T_c$ to 14 mK}{By analyzing the relationship between the $T_c$ of AlMn film and annealing temperature, we optimized the preparation parameters to achieve a $T_c$ in the range of 10 - 20 mK}. Further investigations demonstrated an inverse proportionality between the $T_c$ of AlMn films and applied magnetic field intensity, with lower-$T_c$ specimens exhibiting more pronounced suppression effects. Therefore, magnetic shielding is required for AlMn films in 0$\nu\beta\beta$ experiments. These experimental results provide the critical foundations for ongoing TES detectors fabrication and paves the way for their application in such experiments. Beyond application in 0$\nu\beta\beta$ experiments to clarify the mechanism of neutrinos, ton-scale cryogenic calorimeters with TES detectors can also be employed to detect low-mass dark matter and other particles like coherent neutrino scattering, solar axions, and Majorons. In these applications, AlMn alloy films with extremely low $T_c$ will play important roles. 

\bmhead{Acknowledgements}

This work is supported by the National Key Research and Development Program of China (Grant  No. 2021YFC2203402, No. 2023YFC2206502), the National Natural Science Foundation of China (Grant No. 12141502, No. 12275292).


\bibliography{sn-bibliography}
\end{document}